 \documentclass[aps,prb,preprint,superscriptaddress]{revtex4}
\usepackage{graphicx}

\newcommand\fp{\mbox{$f_{\rm pk}$}}
\def\fig#1{Figure \ref{fig#1}}

\newcommand\resxx{\mbox{Re$( \sigma_{xx} )$}} 
\newcommand\val{\mbox{${v}_{\rm Al}$}}
\newcommand\eal{\mbox{$\mathcal{E}$}}

\newcommand\niddd{ \mbox{$n_{\rm i3d}$}}

\begin{document}



\title{ Pinning modes of high magnetic field Wigner solids with controlled alloy disorder}



\author{B.-H. Moon}
 \affiliation{National High Magnetic Field Laboratory, 1800 E. Paul Dirac Drive, Tallahassee, FL 32310}
 \author{L. W. Engel}
\affiliation{National High Magnetic Field Laboratory, 1800 E. Paul Dirac Drive, Tallahassee, FL 32310}
\author{D. C. Tsui}
\affiliation{Department of Electrical Engineering, Princeton University, Princeton, NJ 08544}
\author{L. N. Pfeiffer}
\affiliation{Department of Electrical Engineering, Princeton University, Princeton, NJ 08544}
\author{K. W. West}
\affiliation{Department of Electrical Engineering, Princeton University, Princeton, NJ 08544}


\date{\today}

\begin{abstract}
  For a  series of samples with  2D electron systems in dilute Al$_x$Ga$_{1-x}$As, with varying $x$ from $0$ to  $0.8\%$, we survey the pinning mode resonances of Wigner solids at the low Landau filling termination of  the fractional quantum Hall effect (FQHE) series.   For all $x$ studied, the   pinning modes  are present with  frequencies, \fp, that  are consistent with collective weak  pinning.   For $x\ge0.22\%$ we find   \fp\ vs $B$  exhibits a rapid increase that is not present for $x=0$.   We   find the observed \fp\  is much smaller than 
  values calculated with a simple Wigner solid model which neglects the effects of the disorder on the charge distribution 
 of a carrier.

\end{abstract}


\maketitle

In two-dimensional electron systems (2DES) hosted in GaAs  an insulating phase  occurs  at  the low Landau filling factor  ($\nu$)  termination of the fractional quantum Hall effect (FQHE) \cite{fqhorig} series.  The insulator  is understood as a form of Wigner solid\cite{wcpredict,jaincfnew,jaincfwc,msreview,db, williams91,willett1,willett2,buhmann,kukushkinwctri,yewc,lessc,murthyrvw},  a  state of matter composed of charge carriers arranged to minimize their mutual repulsion while constrained to a fixed density  by charge neutrality.   The role of the disorder is crucial in producing the insulating behavior, since the Wigner solid is necessarily pinned  by any disorder.   

The spectra of the low $\nu$ insulating phases of 2DES exhibit  a microwave or rf resonance\cite{db,williams91,willett1,willett2,lessc,yewc}   that is  identified as a pinning mode 
\cite{flr,williams91,willett2,lessc,yewc,murthyrvw}, in which pieces of the solid oscillate collectively   about their pinned positions.    This disorder-induced mode is a powerful tool for study of solid phases in 2DES, which occur under many conditions\cite{murthyrvw}  
besides the low $\nu$ insulator.   But  a detailed  understanding  of how the solid phase is pinned is lacking.  
The frequency, \fp, of the resonance   
has been qualitatively found to increase  with increasing accidental disorder, as roughly characterized by the mobilities of a variety of samples, but  there is no systematic experimental knowledge of the relationship between the disorder characteristics  and the pinning mode, and no microscopic picture of  the solid  in the presence of disorder. 

This paper presents pinning mode measurements  of the low $\nu$ insulator in a series of samples with quantifiable, controlled  alloy disorder, produced by    using a dilute Al alloy,  Al$_x$Ga$_{1-x}$As, as the  channel  in which  the 2DES resides.  The characteristics of these samples    have been well established in earlier work on these wafers \cite{wanliscatt,wanlireentrant,wanlisclg,wanlisclganderson,wanlisclgcross}.  In particular, study \cite{wanliscatt} of the mobility vs $x$ has shown that for $x\le 0.85$\% the Al is randomly distributed in the channel, and has allowed modeling the potential around each Al atom as a spherical square well with   known  radius and potential-depth parameters.  Relevant for high magnetic field ($B$) studies, both this radius and the  typical Al spacing in three dimensions are much less than the magnetic length, $l_B$, for all nonzero $x$ that we studied, and for any accessible $B$.    The samples we studied with $x>0$   had  $\sim 10^2 $ to $10^3$ Al within the typical volume of a carrier.

An important aspect of the disorder in 2DES in high $B$ is its effect on the   boundary between the insulating phase and fractional quantum Hall liquid states.    The lowest disorder n-type 2DES  in GaAs are insulating for $\nu$ up to the 1/5 FQHE, and also have a  narrow  reentrant $\nu$ range of insulating phase just above the 1/5 FQHE.   Samples of somewhat larger  disorder remain insulating up to the 1/3 FQHE\cite{shahar,wwkyang}.   One explanation of this is that  the pinning energy, which allows the solid to reduce the charge near high points in the impurity potential,   can stabilize the Wigner solid\cite{pricezhulouie}   against transition to fractional quantum Hall liquids.   Using samples from some of the same wafers studied in this paper, ref \citep{wanlireentrant} describes how the phase boundary between the insulating phase and fractional quantum Hall liquid states is shifted to higher $\nu$  by   dilute Al alloy disorder, and demonstrates  the  terminal (lowest $\nu$) FQHE   is at  1/3  for $x\ge 0.21\%$.   For the samples of  Ref. \onlinecite{wanlireentrant}, and in one other case\cite{wwkyang}, a reentrant range of insulator in a narrow range of $\nu$ above the 1/3 FQHE can be present.

  We find   that the resonance is clearly observable even at the largest $x$ of 0.85\%, at which  \fp\ is increased from its $x=0$ value by about one order of magnitude.  
Considering only the  known densities of the samples, and using a classical shear modulus\cite{bm} the disorder-enhanced \fp\ are  consistent with the weak pinning picture\cite{fl,chitra,fertig,foglerhuse},   in which the carrier positions are determined by a tradeoff between  the disorder potential and the electron-electron interaction, and the disorder potential is not completely dominant.   In the $x>0$ samples \fp\ increases rapidly as $\nu$ decreases into the insulating phase, in rough agreement with the theories.   We estimate the  parameters of the  disorder in a simple model that  takes  a carrier  of the lattice to have the fixed charge distribution of a lowest Landau level orbital.   Using these disorder parameters in   weak pinning theories\cite{chitra,foglerhuse}  overestimates the effect of the   alloy disorder, and predicts that pinning resonances, if present at all, would be at much higher frequency  than we observed.   This discrepancy could be resolved by  the  charge distribution at a carrier in the disordered samples  responding in a way that reduces pinning energy. 



The form of 
transmission-line-based\cite{db}  broadband microwave spectroscopy technique we use has been described elsewhere \cite{zhwuneq,murthyrvw}.   We calculated the real part of diagonal conductivity, \resxx,  from the 
loss on  transmission line patterned in metal film directly on the sample surface, about   180 nm above the 2DES.
The transmission is of coplanar waveguide type, as is  shown schematically in \fig{bsc}a.   We used a  room-temperature
transmitter and receiver  to measure the transmitted power, $P$, and present data 
calculated from $\mbox{Re}(\sigma _{xx})\approx -W|\ln(P/P_{0})|/2Z_0d$  
 where $W=30\ \mu$m is the slot width between center line and ground plane of the coplanar waveguide,  $P_0$ is the reference power  obtained    from an average of transmitted power at $\nu=1$ and $2$.  $Z_0=50\ \Omega$ is the characteristic impedance calculated from the transmission line geometry for $\sigma_{xx}=0$, and $d$ is the length of the line ($d=28$ mm).
The formula is a  low loss, high frequency approximation valid  to within about 20\%, from comparison with a detailed calculation, like that described in Ref.  \onlinecite{zhwuneq},  including the distributed capacitive coupling and allowing for higher loss and reflection.     We estimate the 2DES temperature to be approximately 50 mK, as read by nearby resistance thermometers.   Slight broadening of the resonance could be discerned on increasing the  temperature above that value.   The data are taken in the low microwave power limit, in which further decrease of power (at all frequencies measured, and at 50 mK) does not affect the measurement.  

In the samples we studied the 2DES  is confined at a  single Al$_{x_s}$Ga$_{{1-x}_s}$As /Al$_x$Ga$_{1-x}$As heterojunction, where $x=0,0.21,0.33,0.4,0.8 $ or $0.85 \%$ for the layer in which the 2DES mainly resides, and $x_s\gg x$
is the Al concentration of the spacer layer.    Table \ref{tab1} shows typical densities ($n$) for the microwave measurements, and low temperature mobilities from different pieces of the same wafers\cite{wanliscatt,wanlithesis} , scaled\cite{pfeiffermu} to the densities shown.   The samples were grown in  two series with different characteristics\cite{wanlithesis};  series 1  includes  $x=0,0.21,0.33$ and $0.85\%$ with $x_s=30\%$, and  series 2 includes $x=0.4$ and $0.8\%$ with $x_s=10\%$.  There is other  disorder present in addition to  that due to  the Al  in the channel.
    This background disorder arises from the  potential  from remote ionized donors, alloy disorder in the spacer, interface roughness and accidental impurities introduced throughout the growth;    this background disorder  is what  gives  rise to the pinning mode in the $x=0$ sample. 
    The mobilities in Table 1 clearly show a larger   background disorder for series 1;  for example  the $x=0.4\%$ sample of  series 2 has  higher mobility 
than lower $x$ samples in  the series 1.  Various densities, $n$, were prepared  by brief  low temperature illumination    before the measurements.



\begin{table}[htdp]
\begin{center}
\begin{tabular}{|c|c|c|c|c|}
\hline
$x$   &Wafer &     $\mu $  & $n$  \\
(\%) &Name &    (10$^6$ cm$^2$/V-s)   & ($10^{10}$ cm$^{-2}$)   \\
\hline
 0&7-30-97-2  &   2.0   & 4.6    \\
\hline
 0.21& 8-21-97-2   &   2.3  &   5.9 \\   
\hline
  0.33&8-06-97-1   &    1.7  & 5.6 \\
\hline
0.40 & 12-03-04-1  &    2.6     & 4.9  \\ 
\hline
  0.80 & 12-06-04-1  &   1.6  & 4.9\\
  \hline
  0.85 &  7-30-97-2  &   0.72   &  8.2 \\   
\hline
\end{tabular}
\end{center}
\caption{ Sample parameters, including Al fraction ($x$) for the layer in which the 2DES mainly resides,   mobility ($\mu$)  and  areal density $n$.      }\label{tab1}
\end{table}

\begin{figure}
 \includegraphics[width=3in]{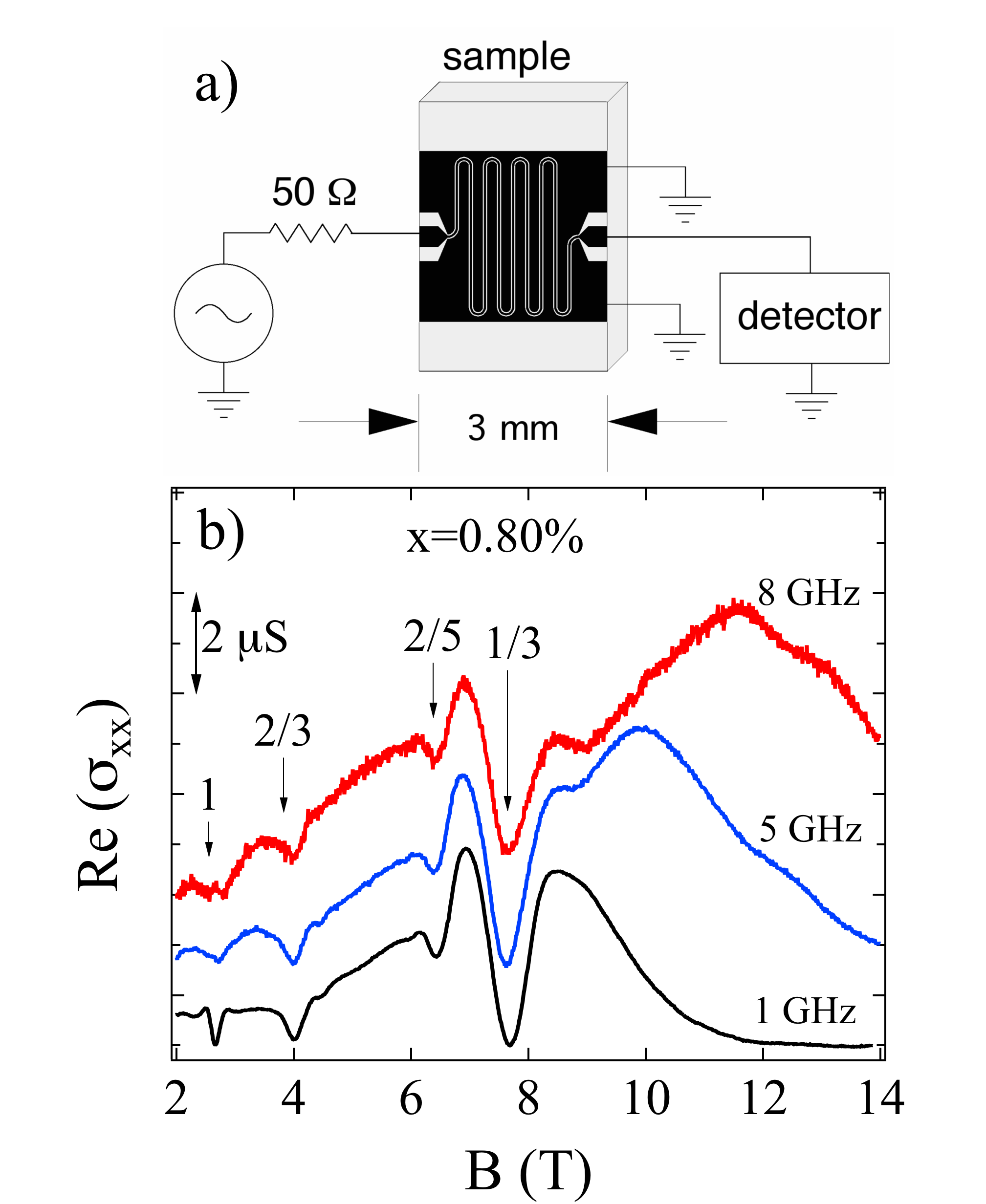}
 \caption{a)  Schematic representation of the microwave circuit used in our measurement, not to scale.    Black areas represent metal films on the sample surface. \ \ b)   Real diagonal conductivity, \resxx, vs magnetic field, $B$, for Al
 fraction $x=0.8 \%$, density $n=6.2 \times 10^{10}$ cm$^{-2}$ at three different frequencies.}\label{figbsc}
 \end{figure}

 \begin{figure*}[t]
 \includegraphics[width=6.5in]{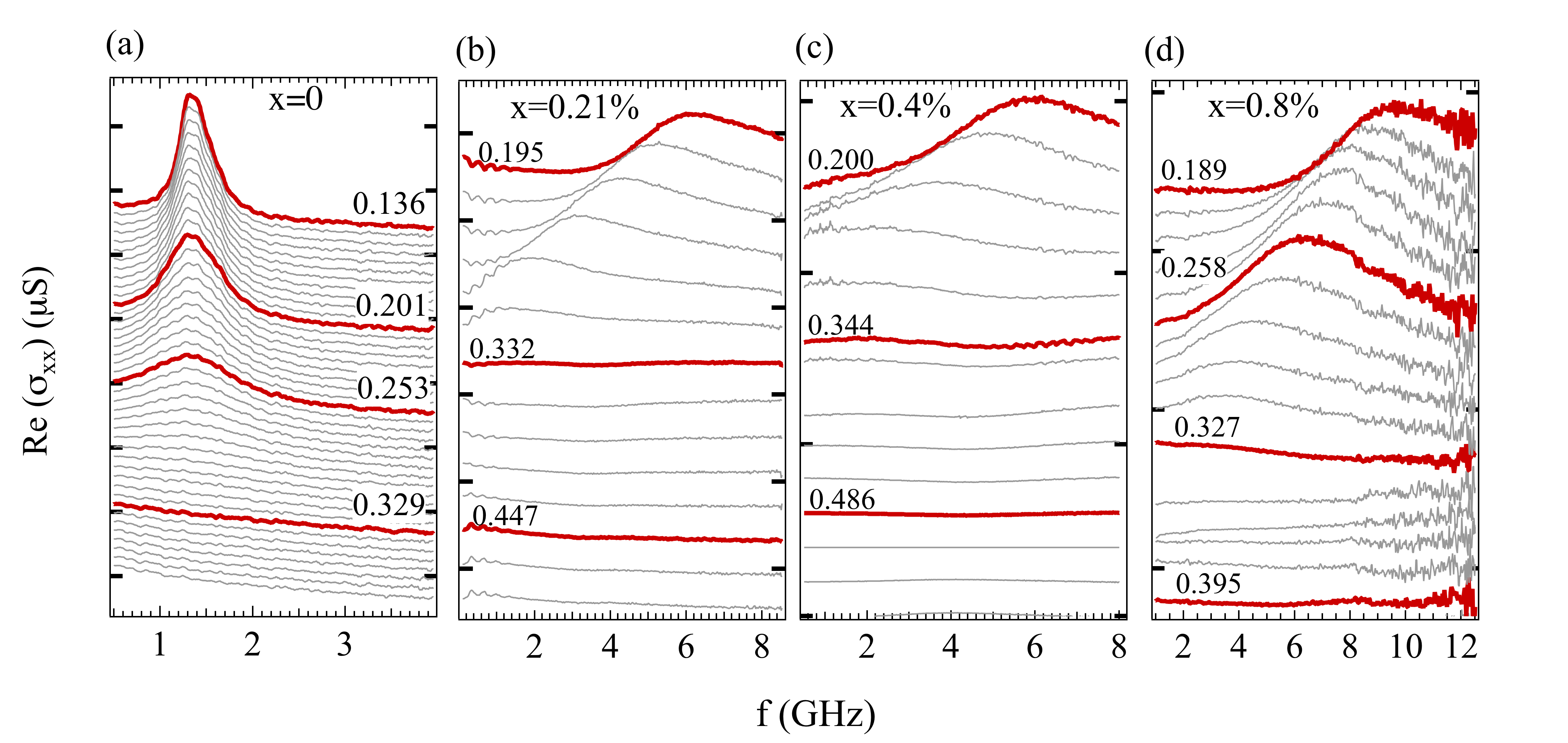}
 \caption{Spectra, real diagonal conductivity \resxx\ vs frequency, $f$, offset vertically for successively lower Landau fillings, $\nu$. $\nu$ for heavy (red online) lines  is marked on graphs. Tick  marks on the vertical axes are separated by 10 $\mu$S.  a) Al fraction $x=0$, density $n=4.6 \times 10^{10}$ cm$^{-2}$. Successive traces were taken at $\nu$ decremented by $\Delta \nu=0.0065$,
 \  \  b) $x=0.21 \%$, $n=8.0 \times 10^{10}$ cm$^{-2}$, $\Delta \nu=0.0058$.
 \ \ c) $x=0.40 \%$, $n=6.7\times 10^{10}$ cm$^{-2}$, $\Delta \nu=0.029$.
 \ \ d) $x=0.80 \%$, $n=6.2 \times 10^{10}$ cm$^{-2}$, $\Delta \nu=0.0137$.}
 \label{figspx}
 \end{figure*}	

\begin{figure}
 \includegraphics[width=3in]{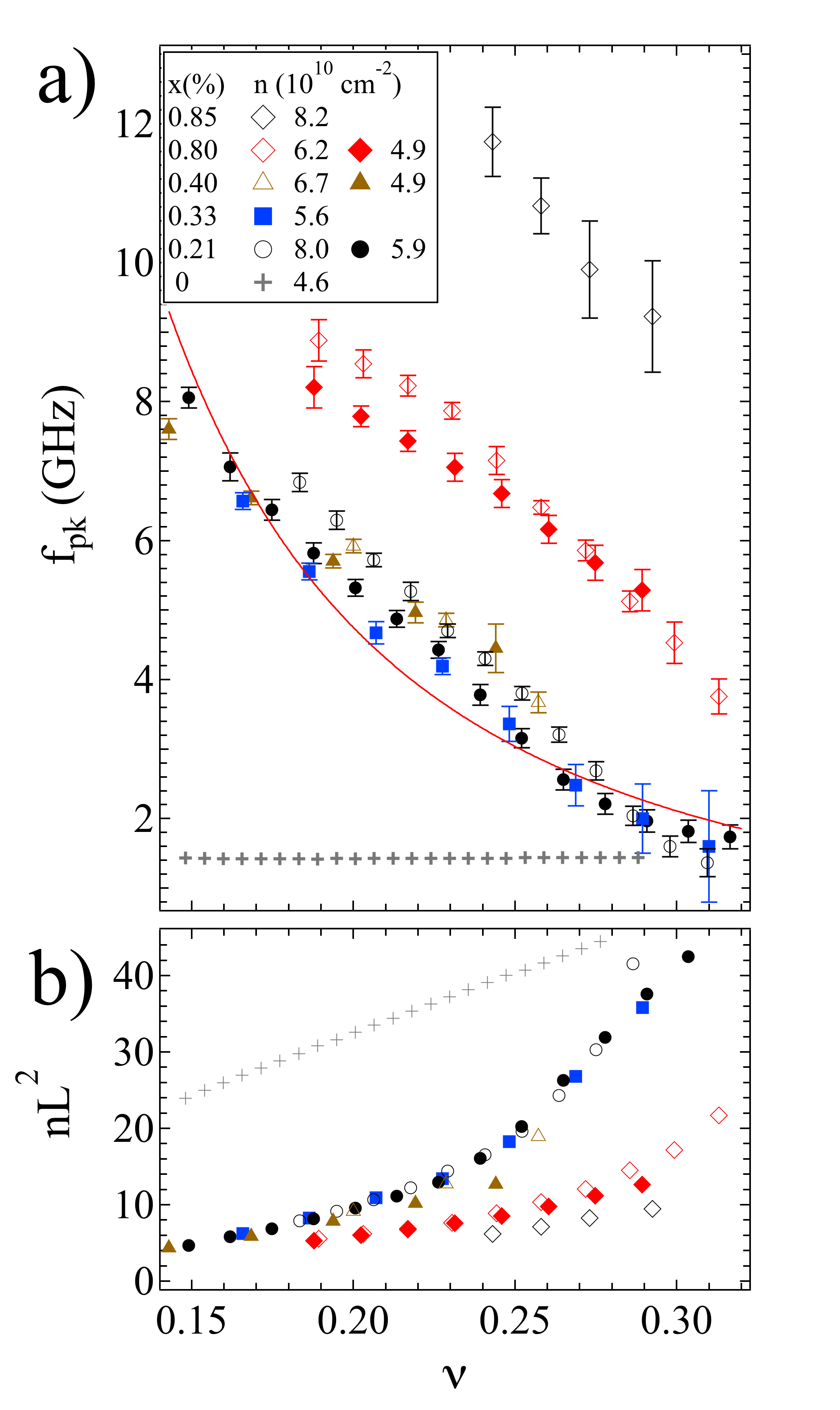}
 \caption{a)  Resonance peak frequency, \fp, vs Landau filling, $\nu$, for various Al alloy fractions $x$ and densities $n$, as marked in the legend.  Thin curve is $\fp=0.19/\nu^2$. b) Number of carriers per domain vs $\nu$, symbols same as in a.  }
   \label{figfp}
 \end{figure}

 \fig{bsc}b shows \resxx\ vs $B$ at $1$,  $5$ and $8$ GHz, for the sample with $x=0.8\%$.    Dips at $\nu= 1/3$ and $2/5$  are clearly visible confirming the finding \cite{wanlireentrant}  that the alloy disorder does not inhibit formation of these FQHE states.  FQHE features are present for the  lower $x$ samples as well.  In the figure there is  slight decrease of $n$, of about 3\% as $B$ increases to 8 T, and the $\nu$ marked on the figure   are taken from the 1/3 FQHE minimum.  $\nu $ is calculated  from that minimum for all data presented in this paper.   As $\nu$ goes below 1/3, the 5 and 8 GHz traces  increase  markedly owing to the onset of the pinning resonance. Reference \onlinecite{wanlireentrant} reported  a reentrant insulating phase for  $\nu$ between the 1/3 and 2/5 FQHE's for an Al alloyed   sample with $x=0.85\%$ and much larger $n$; the data in \fig{bsc}b show no sign of a   resonance   in that $\nu$ range.

\fig{spx} shows  spectra taken at many  $\nu$  for the samples with $x=0,0.21,0.4$ and $0.8\%$.    
For $x=0$, at    $n=4.6 \times 10^{10}$ cm$^{-2}$,  the resonance emerges   as $\nu$ is decreased below 1/3,   growing sharper and developing large peak conductivity   as $\nu$ is decreased further. 
  The resonance appears superposed on a smaller, decreasing smooth background, which is an artifact of the normalization power $P_0$. $P_0$, an average of measurements taken at integer $\nu$,  had a component  that decreased with $f$,  which we ascribe to parallel conduction.    The background  has no significance for our results. 

A resonance is clearly present for the    samples with $x>0$, as shown in \fig{spx}b, c and d.   As in the $x=0$ sample of  \fig{spx}a, the resonance  develops as $\nu$ is decreased below 1/3.   For  all the samples with $x>0$, 
  the resonance frequency shifts upwards as $\nu$ decreases. 


 \fig{fp}a shows  \fp\ vs $\nu$ for all the samples.   \fp\ for $x=0$ can be regarded as a background, and is much smaller than the \fp\ for other curves, except where the resonance is quite weak, near the low-$\nu$ edge of the 1/3 FQHE.   This means  that for the well-developed resonances with  $x>0$, the Al dilute alloy is the main contribution to the pinning frequency.     For $0.2 \le x\le 0.4  \%$  there is 
 little variation in \fp\ vs $\nu$ for the several $n$ values presented.    However, on increasing $x$ to $0.8\%$
 there is again a  definite increase in \fp.    The relatively small change of \fp\  between $x=0.2$ and $0.4 \%$ may be explainable by background disorder,  contributed by other sources than the deliberately introduced Al.    Except for the $  0.21$ and $0.33\%$ samples of    series 1, whose resonance frequencies   are close together,    \fp\ at any $\nu$ increases with  $x$ within each series.

In the collective pinning case,    the solid  deformation  occurs over a large enough length that deformation energy can be calculated from the shear modulus     $\mu_s$. Theories  \cite{fl,chitra, foglerhuse,fertig}  for the collective pinning case connect a 
correlation length of the crystalline order  to \fp\  independent of   other characteristics of   the disorder, as   
$ L=  ( \mu_{s} \pi / 2n eB   \fp )^{1/2}$.
  For collective   pinning to apply, $n L^2\gg  1$ is required.  As a rough estimate we take 
 $\mu_s$ to be the classical value for point particles in a triangular lattice \cite{bm} 
$\mu_{cl}=  0.245 e^2/4\pi\epsilon\epsilon_0 n ^{3/2}$.  Calculations \cite{jaincfnew,narevichfertig,jaincfwc} that take quantum effects into account using the composite fermion  picture give values of $\mu_s$ that are  about 30\% larger than $\mu_{cl}$ 
for the relevant range of $\nu$.
 \fig{fp}b shows the  the number of carriers per domain ($nL^2$)  vs $\nu$, with  $L$  calculated in this way  using  $\mu_{cl}(n)$.     Particularly for larger $\nu$,  including conditions  under which the resonance is well developed, the  collective pinning condition $n L^2\gg  1$ is  satisfied.    

A  semiclassical picture \cite{foglerhuse,chitra,fertig} was developed to explain high magnetic field Wigner solid pinning in weak disorder  with correlation length smaller than $l_B$. In the semiclassical model the carrier guiding centers move  in an ``effective potential"  that results from convolving the true impurity potential with the charge distribution ``form factor" of a carrier.   Using the notation of Ref. \onlinecite{foglerhuse}, the   true impurity potential is characterized  by its   correlation length $\xi$ and its variance $C(0)$.    The charge distribution of each carrier is taken from single particle lowest Landau level orbitals in symmetric gauge, a Gaussian of characteristic radius $l_B$.   
  In the case   $\xi\ll l_B$, which would holds true for the Al alloyed samples considered here,      the  effective  potential (resulting from the convolution)  has correlation length $l_B$ and variance $C(0)\xi^2/l_B^2$.   Minimizing the  total of deformation and pinning energies gives \begin{equation} 
  L=   \frac{\mu_s l_B^3   (2\pi)^{1/2}} {{n^{1/2}}\xi  C(0)^{1/2}}, \ \ \ \  \fp=\frac{C_0\xi^2 } {4 eB \mu_s   l_B^6    }.\label{eq1}\end{equation}.

The  theories  \cite{foglerhuse,chitra} predict that $\fp\propto \nu^{-2}$ for collective pinning with the $\xi\ll l_B$.   A curve of $\fp\propto \nu^{-2}$ is superposed on the data in \fig{fp}a. While the  data   are clearly not a fit to $\fp\propto \nu^{-2}$,   the relation roughly describes the extent of the change of \fp\ over the experimental range of $\nu$.  
The discrepancy between the data and   $\fp\propto \nu^{-2}$ may be due to  an increase $\mu_s$  as $\nu$ decreases.  Such  variation of  $\mu_s$ vs $\nu$  has been  predicted \cite{jaincfnew,narevichfertig}, using  theory based on composite fermion Wigner solids.     
 The   change  of \fp\ with   $\nu$  in \fig{fp}a is in contrast with the essentially flat \fp\ vs $\nu$ for $x=0$, and is an effect of the alloy disorder.  In other work, moderate mobility samples without deliberately induced disorder  \cite{yewc,lessc,clibdep}  have   \fp\   vs $\nu^-1$ increasing    sublinearly as  $\nu$ enters the insulating phase or exhibiting a shallow maximum \cite{yewc}.          

  It is of interest to try to apply the known parameters of the alloy disorder potential to the semiclassical weak pinning calculations. 
   In this picture,  the energy due to each Al in contact with the carrier  is $E_{Al} =  \val\eal/2\pi l_B^2 z_0$,   where    $\val=6.16 \times 10^{-29} {\rm m}^{-3},\eal=1.3\ {\rm eV}$ are the volume and depth  of the spherical square well   from Ref. \onlinecite{wanliscatt}, and $ 2\pi l_B^2 z_0$ is  an effective volume of the carrier.  Thickness $z_0$  is estimated from the Fang-Howard wave function  \cite{afsnote}  at the sample density, $n$.  
Defining \niddd\ as the density of the dilute Al, $\niddd=x/v_{0}$, where $v_0$ is the unit cell volume of the 
Al$_x$Ga$_{1-x}$As, there are an average of $N_{Al}=2\pi l_B^2 z_0  \niddd $ Al per carrier.    
     $N_{Al}  \approx 10^3$  and total   pinning energy per carrier  due to Al  is $ E_{Al} N_{\rm Al} /k_B\approx$ 140 K.  With randomly distributed Al, the  variance,  $V_L^2$, of the   pinning energy of a domain (area $L^2$)  is  $E^2_{Al} N_{Al}nL^2$. 
Energy is minimized by setting $V_L= \mu_s \delta^2$, where  $\delta$ is the displacement required to randomize the carrier pinning energy.       Taking $\delta=l_B$   obtains   equations (1) for $L$ and \fp\ with  $C(0)\xi^2$ substituted by  $\eal^2\val^2\niddd/z_0$.
%

The results of this calculation,   however,  are not consistent with the observation of a microwave-range resonance in the $x>0$ samples.  The calculated  \fp\   is about three orders of magnitude too large for  any of the resonances shown in this paper, and the calculated $L$ would be much less than $l_B$. The pinning energy per  Landau orbital carrier is about half the cyclotron energy for the largest $x$ of $0.8\%$ and $B=14$ T.  It is possible that the carrier wave function is affected by disorder, so that the Landau orbital is a poor approximation.   With random short-range disorder  a more spread-out    charge distribution  would produce weaker pinning.   In  equations 1 this larger effective carrier size  could be modeled by  substituting $l_B$ with a   larger  length.  
 Correlations between carriers, such  those incorporated in the  composite fermion picture \cite{jaincfnew} are also likely to be of importance.     A more complete picture of a disorder pinned state including correlations was considered by  Yannouleas and Landmann \cite{yl}. They found that larger disorder produces pinned  charge-density-wave-like  states  with reduced  charge  modulation amplitude, compared with that of Wigner crystals   in the low disorder case.

    In summary, we have found well-developed microwave resonances in  2DES hosted in dilute Al$_x$Ga$_{1-x}$As, in the solid at the low $\nu$  termination     of the FQHE series,  for $x $  up   to $0.8\%$.     
 Relative to the $x=0$ case, the Al significantly increases \fp\  and also causes a much stronger increase of \fp\  as $\nu$ decreases into the range of the solid phase.  Semiclassical theory for the case of weak disorder, in which a  carrier is taken as having  a lowest Landau level single-particle  form, is not adequate to explain the observation of the  pinning modes in the     Al alloyed samples.


We thank Kun Yang and J. K. Jain for helpful discussions.    The work at Princeton was partially funded by the Gordon and Betty Moore Foundation and the NSF MRSEC Program through the Princeton Center for Complex Materials (DMR-0819860).  
The microwave spectroscopy work at NHMFL was supported  through DOE grant DE-FG02-05-ER46212 at NHMFL/FSU. NHMFL is supported by NSF Cooperative Agreement No. DMR-0084173, the State of Florida and the DOE.
 


\begin{thebibliography}{}
 \bibitem{fqhorig}D. C. Tsui, H. L. Stormer, and A. C. Gossard Phys. Rev. Lett. {\bf 48}, 1559 (1982).
 \bibitem{wcpredict}Y. E. Lozovik and V. I. Yudson, 
  JETP Lett.,  {\bf 22}, 11 (1975);
 P. K. Lam and S. M. Girvin,
Phys. Rev. {\bf B 30}, 473 (1984);
   Kun Yang, F. D. M. Haldane, and E. H. Rezayi
Phys. Rev. B {\bf 64}, 081301 (2001).
\bibitem{jaincfnew}	A. C. Archer, Kwon Park, Jainendra K. Jain, Phys. Rev. Lett., {\bf 111}, 146804 (2013).
\bibitem{jaincfwc}Chia-Chen Chang, Csaba T\"{o}ke, Gun Sang Jeon, and Jainendra K. Jain 
Phys. Rev. B {\bf 73}, 155323 (2006);  Chia-Chen Chang, Gun Sang Jeon, and Jainendra K. Jain 
Phys. Rev. Lett.{\bf 94}, 016809 (2005). 
\bibitem{msreview} M. Shayegan, in Perspectives in Quantum Hall 
Effects, edited by S. Das Sarma and A. Pinczuk (Wiley-Interscience, New York, 1997), p. 343.
 
 
   \bibitem{db}E. Y. Andrei, G. Deville, D. C. Glattli, F. I. B.
Williams. E. Paris, and B. Etienne, ``Observation of a magnetically induced 
Wigner solid,''{ Phys. Rev. Lett.} {\bf 60}, 2765 (1988).
 \bibitem{williams91} F. I. B. Williams, P. A. Wright, R. G. Clark, E. Y. Andrei, G. Deville, D. C. Glattli, O. Probst, B. Etienne, C. Dorin, C. T. Foxon, and J. J. Harris
Phys. Rev. Lett. {\bf 66,} 3285 (1991).
 \bibitem{willett1} M. A. Paalanen, R. L. Willett, R. R. Ruel, P. B.
Littlewood, K. W. West, L. N. Pfeiffer and D. J. Bishop, 
Phys. Rev. B {\bf 45}, 11342 (1992).
  \bibitem{willett2} M. A. Paalanen, R. L. Willett, R. R. Ruel, P. B.
Littlewood, K. W. West, L. N. Pfeiffer and D. J. Bishop, 
\prb{\bf 45}, 13784 (1992).
\bibitem{buhmann}H. Buhmann, W. Joss, K. v. Klitzing, I. V.
Kukushkin, A. S. Plaut, G. Martinez, K. Ploog, and V. B. Timofeev, 
 Phys. Rev. Lett. {\bf 66},  926 (1991).

\bibitem{kukushkinwctri}I. V. Kukushkin, Vladimir I. FalÕko, R. J. Haug, K. von Klitzing, K. Eberl, and K. Tštemayer
Phys. Rev. Lett. 72, 3594(1994).
\bibitem{yewc} P. D. Ye, L. W. Engel, D. C. Tsui, R. M. Lewis, L. N. Pfeiffer, and K. West, Phys. Rev. Lett. {\bf 89}, 176802 (2002).

%
 \bibitem{lessc}  L. W. Engel ,C.-C. Li, D. Shahar, D. C. Tsui and 
 M. Shayegan,   
 Solid State Commun., {\bf 104} 167-171  (1997).

 \bibitem{murthyrvw}  G. Sambandamurthy,  Zhihai Wang,  R.M. Lewis, Yong P. Chen, L.W. Engel,
D.C. Tsui, L.N. Pfeiffer  and  K.W. West,
 Solid State Commun. {\bf 140}, 100  (2006)  contains a brief review. 

\bibitem{flr}Hidetoshi Fukuyama and Patrick A. Lee
Phys. Rev. B {\bf 18}, 6245  (1978).



 



 \bibitem{wanliscatt}Wanli Li, G.A. Cs\'{a}thy, D.C. Tsui, L.N. Pfeiffer, and K.W. West, Appl. Phys. Lett. 83, 2832 (2003).
 \bibitem{wanlireentrant} Wanli Li, D. R. Luhman, D. C. Tsui, L. N. Pfeiffer, and K. W. West
Phys. Rev. Lett. {\bf 105}, 076803 (2010).

\bibitem{wanlisclg}
Wanli Li, G.A. Cs\'{a}thy, D. C. Tsui, L. N. Pfeiffer, and K. W. West
Phys. Rev. Lett. {\bf 94}, 206807 (2005).
\bibitem{wanlisclganderson}
Wanli Li, C. L. Vicente, J. S. Xia, W. Pan, D. C. Tsui, L. N. Pfeiffer, and K. W. West
Phys. Rev. Lett. {\bf 102}, 216801 (2009).
\bibitem{wanlisclgcross}
Wanli Li, J. S. Xia, C. Vicente, N. S. Sullivan, W. Pan, D. C. Tsui, L. N. Pfeiffer, and K. W. West
Phys. Rev. B {\bf 81}, 033305 (2010).
 
 
 \bibitem{shahar}D. Shahar, D. C. Tsui, M. Shayegan,   R. N. Bhatt and 
 J. E. Cunningham, Phys. Rev. Lett. {\bf 74}, 4511 (1995).
\bibitem{wwkyang} I. Yang, W. Kang, S. T. Hannahs, L.N. Pfeiffer, and K.W. West, Phys. Rev. B {\bf 68}, 121302(R) (2003).

%
\bibitem{pricezhulouie}R. Price, Xuejun Zhu, P. M. Platzman, and Steven G. Louie
Phys. Rev. B {\bf 48}, 11473 (1993).


\bibitem{bm}L. Bonsall and A. A. Maradudin
Phys. Rev. B{\bf 15}, 1959 (1977).
\bibitem{fl} H. Fukuyama, and P. A. Lee, Phys. Rev. B  {\bf 18}, 6245 (1978).
 
  \bibitem{chitra} R. Chitra, T. Giamarchi, and P. Le Doussal, 
Phys. Rev. Lett. 80, 3827 (1998); R. Chitra, T. Giamarchi, and P. Le Doussal,  Phys. Rev. B {\bf 65}, 035312 (2001).
 \bibitem{fertig} H. A. Fertig, Phys. Rev. B {\bf 59}, 2120 (1999).
\bibitem{foglerhuse} M. M. Fogler, and D. A. Huse, Phys. Rev. B  {\bf 62}, 7553 (2000).

\bibitem{zhwuneq}Zhihai Wang, Yong P. Chen, Han Zhu, L. W. Engel, D. C. Tsui, E. Tutuc, and M. Shayegan
Phys. Rev. B {\bf 85}, 195408 (2012).


\bibitem{wanlithesis}Wanli Li, doctoral dissertation, Princeton University, 2007. 
  \bibitem{pfeiffermu}Loren Pfeiffer, K. W. West, H. L. Stormer, and K. W. Baldwin
 Appl. Phys. Lett. {\bf 55}, 1888 (1989).
%
\bibitem{narevichfertig}R. Narevich, Ganpathy Murthy, and H. A. Fertig
Phys. Rev. B {\bf 64}, 245326 (2001).
 
%
 
 \bibitem{clibdep}C.-C. Li, L. W. Engel, D. Shahar, D. C. Tsui, and M. Shayegan
Phys. Rev. Lett.  {\bf 79}, 1353(1997).


 \bibitem{afsnote}F.F. Fang and W. E. Howard, Phys. Rev. Lett. {\bf16}, 797 (1966).  We use
  $z_0=3 (8\hbar^2\epsilon\epsilon_0/33m^* e^2 n)^{1/3}$, where $m^*$ is the band mass.   
  For the 2DES studied here, $n$ was between $4.9$ and $8\times 10^{10}$ cm$^{-2}$, from which  the calculated $z_0$   was between 22 and  19 nm. 
  
  
 \bibitem{yl}
C. Yannouleas and U. Landman, Phys. Rev. B {\bf 84}, 165327 (2011).
 
 
 
\end{thebibliography}
\end{document}